\documentclass[superscriptaddress,longbibliography,floatfix,twocolumn,nobibnotes]{revtex4-2}
\usepackage{upgreek}
\usepackage{amsmath,braket}
\usepackage{graphicx}
\usepackage{mathtools}
\usepackage{color}
\usepackage{xcolor}
\usepackage{setspace}
\newcommand{\RNum}[1]{\uppercase\expandafter{\romannumeral #1\relax}}
\usepackage{comment}
\usepackage{soul}


\begin{document}
\title{Cavity-enhanced excitation of a quantum dot in the picosecond regime}

\author{Alisa Javadi}
\email{alisa.javadi@unibas.ch}
\affiliation{Department of Physics, University of Basel, Klingelbergstrasse 82, CH-4056 Basel, Switzerland}

\author{Natasha Tomm}
\affiliation{Department of Physics, University of Basel, Klingelbergstrasse 82, CH-4056 Basel, Switzerland}

\author{Nadia O. Antoniadis}
\affiliation{Department of Physics, University of Basel, Klingelbergstrasse 82, CH-4056 Basel, Switzerland}

\author{Alistair J. Brash}
\affiliation{Department of Physics and Astronomy, University of Sheffield, Sheffield S3 7RH, United Kingdom}

\author{R\"{u}diger Schott}
\affiliation{Lehrstuhl f\"{u}r Angewandte Festk\"{o}rperphysik, Ruhr-Universit\"{a}t Bochum, D-44780 Bochum, Germany}

\author{Sascha R. Valentin}
\affiliation{Lehrstuhl f\"{u}r Angewandte Festk\"{o}rperphysik, Ruhr-Universit\"{a}t Bochum, D-44780 Bochum, Germany}

\author{Andreas D. Wieck}
\affiliation{Lehrstuhl f\"{u}r Angewandte Festk\"{o}rperphysik, Ruhr-Universit\"{a}t Bochum, D-44780 Bochum, Germany}

\author{Arne Ludwig}
\affiliation{Lehrstuhl f\"{u}r Angewandte Festk\"{o}rperphysik, Ruhr-Universit\"{a}t Bochum, D-44780 Bochum, Germany}

\author{Richard J. Warburton}
\affiliation{Department of Physics, University of Basel, Klingelbergstrasse 82, CH-4056 Basel, Switzerland}

\date{\today}

\begin{abstract}
{A major challenge in generating single photons with a single emitter is to excite the emitter while avoiding laser leakage into the collection path. Ideally, any scheme to suppress this leakage should  not result in a loss in efficiency of the single-photon source. Here, we investigate a scheme in which a single emitter, a semiconductor quantum dot, is embedded in a microcavity. The scheme exploits the splitting of the cavity mode into two orthogonally-polarised modes: one mode is used for excitation, the other for collection. By linking experiment to theory, we show that the best population inversion is achieved with a laser pulse detuned from the quantum emitter. The Rabi oscillations have an unusual dependence on pulse power. Our theory describes them quantitatively allowing us to determine the absolute photon creation probability. For the optimal laser detuning, the population innversion is 98\%. The Rabi oscillations depend on the sign of the laser-pulse detuning. We show that this arises from the non-trivial effect of phonons on the exciton dynamics. The exciton-phonon interaction is included in the theory and gives excellent agreement with all the experimental results.}
\end{abstract}

\maketitle

\normalsize

\section{Introduction}
Quantum emitters efficiently interfaced with optical cavities represent primary components in photonic quantum technologies. They are used for generating quantum states of light such as single photons and entangled states. The generation efficiency requirement is strict, with many proposals requiring efficiencies higher than 90\% \cite{GimenoPRL2015,LOPRL2020}. Generating photonic quantum states requires coherent control over the quantum emitter, which is often carried out using fast laser pulses. 
The main challenges are to ensure that the laser pulse results in occupation of the upper level with near-unity probability and that laser light does not enter the collection mode. Another major complication for solid-state quantum emitters is the interaction with the environmental degrees of freedom, in particular the acoustic phonons \cite{Vagov2007PRL,Ramsay2010PRL}.

Several approaches have been developed to separate the excitation pulse from the generated photons. One method excites the emitter via non-cavity modes with a propagation direction perpendicular to the cavity axis, a scheme which is often used to generate photons from atoms and ions \cite{SchuppPRXQ2021,MeranerPRA2020}. In the solid-state domain, non-resonant excitation schemes, such as a phonon-assisted mechanism \cite{Ates2009NPHOT,Thomas2021,Gustin2020,Cosacchi2019PRL}, allow spectral filtering of the laser pulse. However, the essential spectral filtering unavoidably reduces the efficiency of the source. Additionally, phonon-assisted schemes require large pulse areas \cite{Thomas2021}. The pulse area can be reduced to the minimum, $\pi$, by exciting the quantum emitter resonantly. In such  schemes, the collection and excitation modes have a different  spatial \cite{HuberOptica2020} or polarization degree-of-freedom \cite{KuhlmannRSI2013}. It is challenging to avoid losses. For instance, the cross-polarized scheme can result in the loss of 50\% of the generated photons.

In many cases, the cavity mode splits into two modes with orthogonal polarization, a consequence of weak birefringence either in the mirrors or in the solid-state host. This mode structure offers a solution to the excitation-collection challenge. One cavity-mode, resonant with the quantum emitter, is used for collection; the other cavity-mode is used for excitation \cite{Wang2019}. Furthermore, if the quantum emitter has a circularly-polarized optical dipole-moment, a cross-polarized detection scheme does not compromise the efficiency of the source \cite{Wang2019}. The scheme was originally developed for quantum dots (QDs) in semiconductor micropillars, for which the cavity mode-splitting is induced by an elliptical pillar cross-section \cite{Wang2019}. It was subsequently employed in a QD-in-open-cavity device  \cite{Tomm2021}. In this case, the mode-splitting arises from birefringence in the semiconductor heterostructure, and it can be tuned via the electro-optic \cite{Frey2018} and photo-elastic effects \cite{Tomm2021PRA}.

Here, we probe both experimentally and theoretically the cavity-based excitation-collection scheme using a QD coupled to a one-sided open-microcavity, Fig.\,\ref{fig:concept}a. The experiment explores the dependence on both laser detuning and pulse power. The theoretical model describes the effect of the cavity on the excitation pulse. It also includes the exciton-phonon interaction. The model describes the experimental results precisely allowing us to quantify the photon creation probability, to understand the unusual Rabi oscillations, and to predict the behaviour as a function of cavity-mode splitting.

\section{Cavity-mediated excitation of a two-level system}

\begin{figure}[t!]
    \includegraphics[width=\columnwidth]{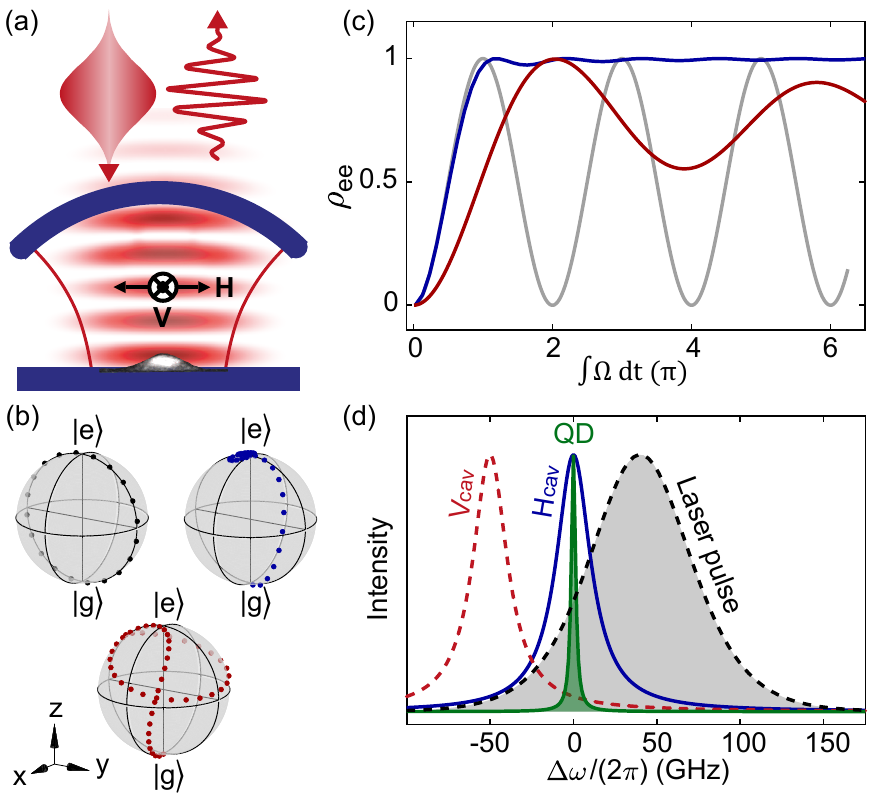}
 	\caption{Excitation mechanisms. (a) Schematic picture of a quantum emitter coupled to a one-sided cavity. The cavity's fundamental mode is split into two non-degenerate H- and V-polarized modes. (b) Bloch-sphere representation of the TLS state when interacting with a resonant Gaussian pulse (black dots), a chirped pulse (blue dots), and a cavity-filtered pulse (red dots), as a function of the pulse area. (c) The excited state population, $\rho_{ee}$, versus pulse area for excitation with a Gaussian pulse (gray line), a chirped pulse (blue line), and a cavity-filtered pulse (red line). (d) Spectral configuration for the cavity-mediated excitation. The quantum emitter is resonantly coupled to the higher-frequency H-polarized cavity mode (blue), and the V-polarized laser pulse interacts with the quantum emitter via the lower-frequency V-polarized cavity mode (red).}
	\label{fig:concept}
\end{figure}

We consider initially pulsed excitation of a two-level system (TLS) in the ideal limit (absence of decay and dephasing processes), where the pulse duration is significantly shorter than the lifetime of the emitter. In the case of a transform-limited pulse, a resonant pulse drives the TLS around the Bloch-sphere, inverting its population from the ground state $\ket{g}$ to the excited state $\ket{e}$, black dots in Fig.\,\ref{fig:concept}b. On increasing the pulse area, the population rotates coherently around the Bloch-sphere, resulting in the well-known Rabi-oscillations, gray line in Fig.\,\ref{fig:concept}c. 

Another well-known case is the dynamics of a TLS under excitation with a strongly chirped pulse, rapid adiabatic passage \cite{Wei2014,Wu2011}. The TLS interacts with different frequency components present in the pulse at different instants in time, resulting in a different trajectory on the Bloch-sphere, blue dots Fig.\,\ref{fig:concept}b. Starting at the south pole of the Bloch sphere, the state tends to gravitate to the north pole, making the excited state population insensitive to the pulse area, Fig.\,\ref{fig:concept}c.

In the case of a cavity-mediated excitation with a detuned pulse, a Gaussian-shaped pulse is convoluted with the time-response of the cavity itself. Effectively, the cavity acts as a dispersive filter, altering the spectral profile of the pulse such that it can no longer be described by a Gaussian profile in the frequency domain. Figure \ref{fig:concept}d shows the spectral configuration of the original laser pulse and the cavity modes. The TLS is resonant with the higher-frequency cavity mode (H-polarized), and the laser pulse is launched via the lower-frequency (V-polarized) cavity mode. The red curves in Fig.\,\ref{fig:concept}b,c show the evolution of the TLS as a function of the input pulse area. A full population in the excited state can be obtained, but depletion of the population is incomplete at higher pulse areas. At high pulse areas, the excited state population converges to a constant value lying between 0 and 100\% dependending on the detuning. This behaviour at high pulse areas is quite different to the other two excitation mechanisms.

\section{Experimental Results\label{sec:experiment}}
\begin{figure}[t!]
        \includegraphics[width=\columnwidth]{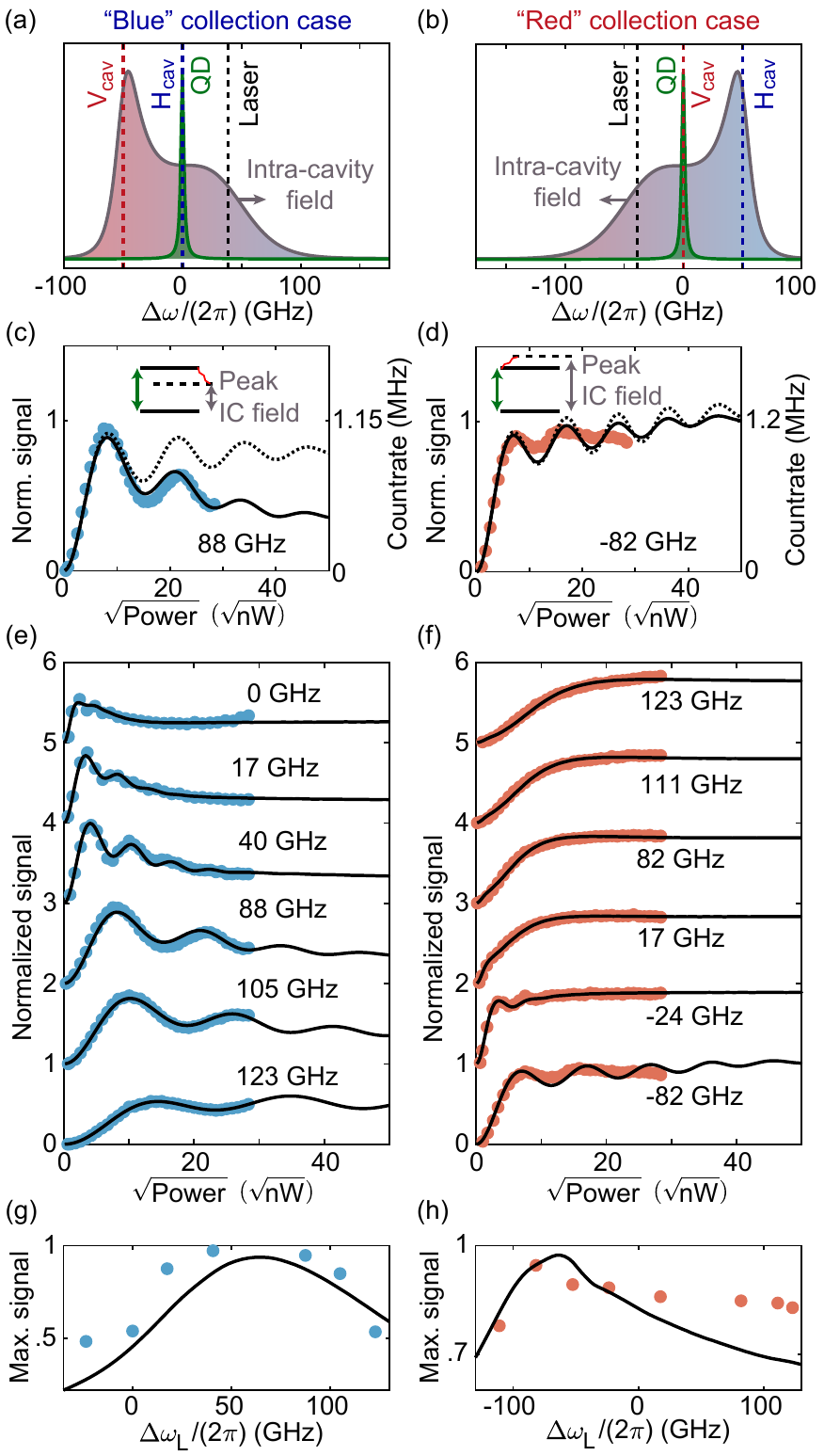}
 	\caption{Normalized photon rate as a function of laser detuning and choice of collection cavity. (a) Schematic of the ``blue" collection case in which the QD is resonant with the higher-frequency cavity-mode and the lower-frequency cavity acts as the excitation cavity. (b) Schematic of the ``red" collection case in which the role of the two cavity-modes is reversed. (c) Normalized photon rate (measured signal normalized to the known losses of the system) as a function of the square-root of the input power for the ``blue" case with $\Delta\omega_{\rm L}/(2\pi)=88$\,GHz. The solid black line is the result for the population inversion in the model including phonons; the dotted line is the result of the model in the absence of phonons. The inset depicts the energy levels of the QD and the peak photon-energy of the intra-cavity field (IC). (d) Same as (c), but for the ``red" case. (e) and (f) Normalized photon rate for ``blue" and ``red" cases, respectively. The data sets are offset by one unit for better visualization. (g) and (h) Measured peak signal and calculated peak population inversion  as a function of laser detuning. The model uses a fixed $t_p=3.6$\,ps.}
	\label{fig:Pi_res}
\end{figure}

We use a QD in an open microcavity \cite{Tomm2021} to study the cavity-mediated excitation scheme. The cavity hosts two modes with orthogonal polarization and a mode-splitting of 50\,GHz. The laser pulses have a temporal width (intensity full-width at half maximum) between 3.6 and 5.0\,ps. We use the polarization of the excitation laser to select the excitation cavity mode. Figure\,\ref{fig:concept}d shows the case where the QD is on resonance with the higher-frequency (H-polarized) mode and the laser is launched through the lower-frequency (V-polarized) mode, the ``blue" collection case. Figure\,\ref{fig:Pi_res}a shows the calculated intra-cavity field in this configuration. As evident from the spectrum, the intra-cavity field has a strong peak at frequencies below the resonance of the QD. Figure\,\ref{fig:Pi_res}b shows the spectrum in the inverted case, the ``red" collection case: the QD is on resonance with the lower-frequency (V-polarized) mode and the laser is launched through the higher-frequency (H-polarized) mode.

We present first experimental results on the ``blue" collection case. Figure\,\ref{fig:Pi_res}c shows the normalized single-photon rate as a function of the input laser power. The laser is detuned from the QD by $\Delta\omega_{\rm L}/(2\pi)=88$\,GHz. We observe the expected oscillatory behaviour along with strong damping as the laser power is increased. The damping is a result of the interaction between the QD exciton, the red-detuned components of the intra-cavity field, and the phonons in the environment of the QD. In particular, the damping can be depicted as the process in which the QD decays by emitting a photon on resonance with the intra-cavity peak and a phonon, depicted in the inset of Fig.\,\ref{fig:Pi_res}c. The process is enhanced by the large amplitude of the intra-cavity field, and hence we observe stronger decays for increasing laser powers. Processes of this nature have already been observed in pump-probe experiments \cite{LiuPRB2016}. The theory (Sec.\,\ref{sec:theory}) captures all these details, shown by the solid lines in Fig.\,\ref{fig:Pi_res}c.

A crucial metric is the population inversion, the probability of creating an exciton in the QD following pulsed excitation. The probability of creating a photon in the collection cavity-mode is $\eta_{c}=\beta_{c} \pi_{e}$ where $\pi_{e}$ stands for the population inversion, and $\beta_{c}$ the probability that an exciton creates a photon in the collection cavity-mode. $\beta_{c}$ (and the other factors which determine the exact measured photon flux \cite{Tomm2021}) remains constant as a function of power. Hence, the convincing match of the theory to the experimental results allows us to deduce that we achieve a maximum population inversion of $\pi_{e}=96$\% in this experiment. The theory also allows us to quantify the exact role of phonons -- the dashed line in Fig.\,\ref{fig:Pi_res}c shows the theory with the exciton-phonon interaction turned off. Phonons limit only slightly the population inversion (99\% $\rightarrow$ 96\%) at small powers but have a significant effect at large powers.

We turn to the ``red" collection case, a scheme mirrored in frequency with respect to the ``blue" case. Figure\,\ref{fig:Pi_res}d shows the normalized single-photon rate as a function of the input laser power with $\Delta\omega_{\rm L}/(2\pi)=-82$\,GHz. Interestingly, the strong damping disappears in this case. This is due to the fact that a phonon-mediated process is suppressed: the peak of the intra-cavity field lies at a higher frequency than the QD resonance, such that phonon-mediated depopulation of the excited state would require absorption of a phonon which is suppressed at 4.2\,K, the temperature of the experiment, due to the low thermal population of the phonon bath. The symmetry breaking in the system's evolution, ``red" with respect to ``blue", is strong evidence for the role of phonons in the system dynamics \cite{KaldeweyPRB2017,Quilter2015PRL}. As for the ``blue" case, the theory reproduces the experimental results very convincingly. 

\begin{figure*}[t!]
    \centering
        \includegraphics[width=\textwidth]{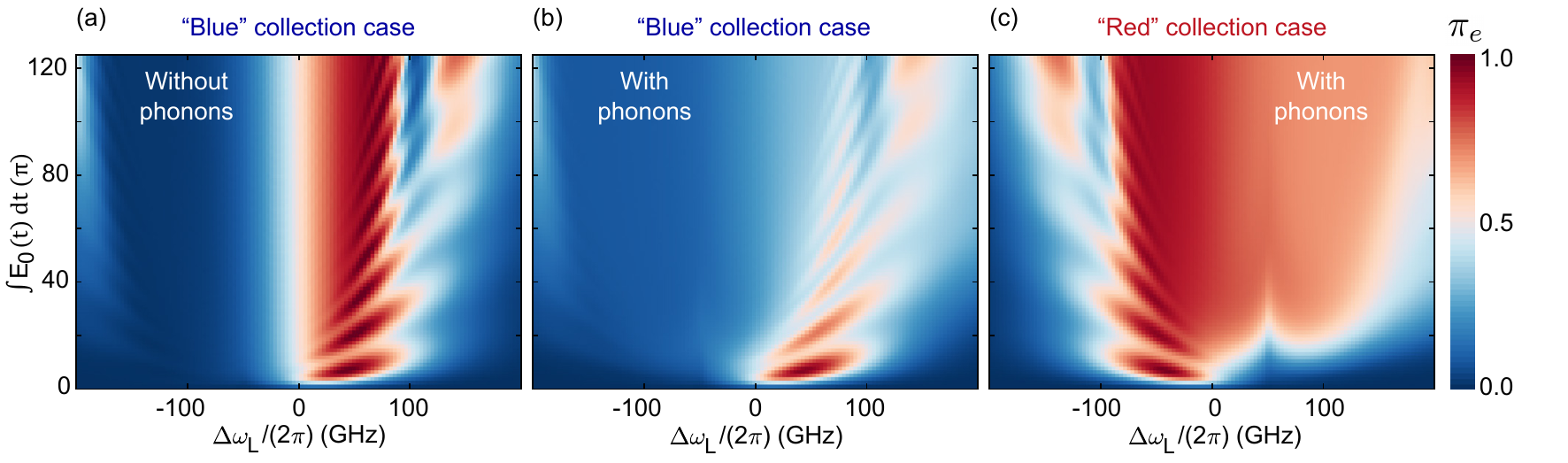}
 	\caption{Simulated population inversion of a QD excited by a cavity-filtered light pulse. 
 	 $\eta_c$ is plotted as a function of input laser detuning from the QD ($\Delta\omega_\text{L}$) and the excitation pulse amplitude. For this simulation, $\kappa/(2\pi) = 25$\,GHz, the splitting between the two orthogonal cavity modes is 50\,GHz, and the input pulse width is $t_p$ = 4.2\,ps. (a) The ``blue" collection case in the absence of phonons. The ``red" collection case is equivalent in the absence of phonons but with symmetrically reflected laser frequencies. (b), (c) The calculated values with the same parameters in the presence of phonons: (b) ``blue" collection case, where $\pi_{e}$ reaches 96.1\%; and (c) the ``red" collection case, where $\pi_{e}$ reaches 97.8\%.}
	\label{fig:simulations}
\end{figure*}

We now investigate the dependence on the laser detuning. Zero detuning corresponds to the case when the laser and the QD are in resonance. Figures \ref{fig:Pi_res}e and f show the measured normalized photon rates for different laser detunings ($\Delta\omega_{\rm L}/(2\pi)$) for the ``blue" and ``red" collection cases, respectively. For the ``blue" collection case (Fig.\,\ref{fig:Pi_res}e), one observes the fingerprint of Rabi rotations along with phonon-induced damping. The peak population inversion is higher than 90\% for $\Delta\omega_{\rm L}/(2\pi)$ between 40\,GHz and 100\,GHz. At resonance and for negative detunings, $\Delta\omega_{\rm L}\leq0$, the peak population inversion decreases drastically. For the ``red" collection case (Fig.\,\ref{fig:Pi_res}f), the oscillatory behaviour is less pronounced and only present when the laser is red-detuned, $\Delta\omega_{\rm L}\leq0$. Interestingly, the population inversion can still be close to unity for $\Delta\omega_{\rm L}\geq0$. This can be described by a phonon-assisted excitation of the QD: in this scenario, the intra-cavity field lies at higher frequencies than the QD transition such that a laser photon can be converted to a QD-exciton and a phonon \cite{Ates2009NPHOT,Quilter2015PRL}. 

Figures \ref{fig:Pi_res}g and \ref{fig:Pi_res}h show the highest measured photon rates as a function of the excitation frequency for the ``red" and ``blue" schemes (data in Figures \ref{fig:Pi_res}e and \ref{fig:Pi_res}f, respectively). In both cases, the maximum is achieved for a detuned pulse.

\section{Theoretical treatment \label{sec:theory}}
We describe the interaction between a TLS coupled to the collection cavity and a driving electric field with the standard Hamiltonian:
\begin{equation}
\begin{split}
    \label{eq:hamiltonian}
    \hat{H} =&\hbar\, \Delta\omega_c\, \hat{a}^\dagger_{c} \hat{a}_{c} + \hbar g \left(\hat{a}^\dagger_{c}\hat{\sigma}_{-} + \hat{a}_{c}\hat{\sigma}_{+}\right)\\
    &+ \hbar \left(\overline{E(t)}\cdot\overline{\mu}\right)\left(e^{i\omega_0 t}\hat{\sigma}_{+} + e^{-i\omega_0 t}\hat{\sigma}_{-}\right). 
\end{split}
\end{equation}
The Hamiltonian is in the rotating frame of the TLS. $\omega_0/(2\pi)$ is the resonance frequency of the TLS, $\Delta\omega_c/(2\pi)$ is the frequency detuning between the collection cavity-mode and the TLS ($\Delta\omega_c=\omega_c-\omega_0$), and $\hat{a}_{c}^\dagger$ is the photon creation operator for the collection cavity. $\overline{E(t)}$ is the intra-cavity field driving the TLS. The leakage through the top mirror can be modeled using the Lindblad operator $\hat{\mathcal{L}}=\sqrt{\kappa}\hat{a_\textrm{c}}$. 

The laser interacts with the TLS via the second cavity mode, the excitation mode, with resonance frequency $\omega_{e}/(2\pi)$. The intra-cavity field is a convolution of the input pulse and the impulse response of the cavity mode. The impulse response of a cavity is $h(\tau)=e^{-\frac{\kappa \tau}{2}}\textrm{cos}(\omega_e \tau)\Theta(\tau)$, where $\kappa/(2\pi)$ is the linewidth of the cavity mode. Assuming an input field $E_0(t)= (\pi t_p)^{-1}\textrm{sech}(t/t_p)\textrm{cos}(\omega_{L}t)$, typical of a mode-locked laser, the intra-cavity field is:
\begin{equation}
\begin{split}
    \label{eq:puls_shape}
    \overline{E(t)}=&\frac{\kappa e^{i\omega_Lt}\textrm{sech}(t/t_p)}{2\pi j_m}\\
    &\times{}_2F_1\left(1,1,1+j_m/2,\textrm{sech}(t/t_p)/2\right)+\textrm{c.c.},\\
\end{split}
\end{equation}
where $j_m=1-(i\Delta\omega_\textrm{EL}-\kappa/2)t_p$, $\omega_L/(2\pi)$ is the centre frequency of the laser, $t_p$ the input pulse width, $\Delta\omega_\textrm{EL}/(2\pi)$ the detuning between the laser and the excitation cavity mode, and ${}_2F_1$ Gauss's hypergeometric function. We plug Eq.\,\ref{eq:puls_shape} into Eq.\,\ref{eq:hamiltonian} and apply the standard rotating-wave approximation.

Phonons play a significant role in the dynamics of a QD, as the instantaneous Rabi frequency can be as high as several terahertz at which the exciton-phonon coupling is strongest. Assuming weak coupling of the exciton to the environment, one can include the effect of phonons on the TLS dynamics using the Bloch-Redfield master equation \cite{Redfield1965,Breuer&Petruccione2002,Lennart2014PRB}. 

We use the Python package Qutip \cite{JOHANSSON2012CMPT,JOHANSSON2013CMPT} to set up and solve the equations of motion based on the Hamiltonian in Eq.\,\ref{eq:hamiltonian}. Finally, the photon creation probability in the collection mode ($\eta_\textrm{c}$) is calculated as $ \beta_\textrm{c} \pi_{e}$ with $\pi_{e}=\int\kappa\left\langle \hat{a}^\dag_\text{c} \hat{a}_\text{c}\right\rangle\, dt$ and $\beta_\textrm{c} = \frac{F_p(\omega_c)}{F_p(\omega_c)+F_p(\omega_e)+1}$, the probability of the QD exciton creating a photon in the collection cavity mode. We note that $\int\kappa\left\langle \hat{a}^\dag_\text{c} \hat{a}_\text{c}\right\rangle\, dt$ is the number of photons generated by the excitation pulse and can in principle can exceed one (via multiple excitations of the TLS by one pulse). However, in the regime explored here (pulse duration much less than radiative decay time), $\int\kappa\left\langle \hat{a}^\dag_\text{c} \hat{a}_\text{c}\right\rangle\, dt$ follows the population of the TLS upper-state very closely and we chose to describe it with ``population inversion".

To extract numerical results, we use the exciton-phonon coupling described in Ref.\,\cite{nazir2016modelling}. We assume a spherically symmetric wavefunction for both electrons and holes ($\psi\propto e^{-r^2/{r_0}^2}$). Our model matches the experimental data very well for an electron radius of 5.9\,nm and a hole radius of 3.6\,nm. We also use $\kappa/(2\pi)=25$\,GHz and a mode-splitting of 50\,GHz extracted from earlier measurements \cite{Tomm2021}. We take a pulse width between 3.6\,ps and 5.0\,ps.

We use our model to map out $\pi_{e}$ as a function of the laser detuning. We use the same parameters from modeling the data in Fig.\,\ref{fig:Pi_res}. Figure\,\ref{fig:simulations}a shows the behaviour of an ideal (phonon-less) QD in the ``blue" collection case. Near-unity population inversion is observed over a range of laser detunings. For a phonon-less QD coupled to the red-detuned cavity mode (``red" collection case), this plot would be mirrored with respect to $\Delta\omega_{\rm L} = 0$. Figures\,\ref{fig:simulations}b and c show the population inversion including the interaction with phonons. The effect of phonons is visible in the striking difference between these two plots and Fig.\,\ref{fig:simulations}a. Notably, despite the asymmetric behaviour, ``blue" with respect to ``red" collection modes, these results clearly show that near-unity population inversion can be obtained in both cases. These results also clearly demonstrate that the maximum population inversion is not obtained at a strict resonant condition when exploiting the cavity-mediated excitation scheme.

\begin{figure}[ht!]
        \includegraphics[width=\columnwidth]{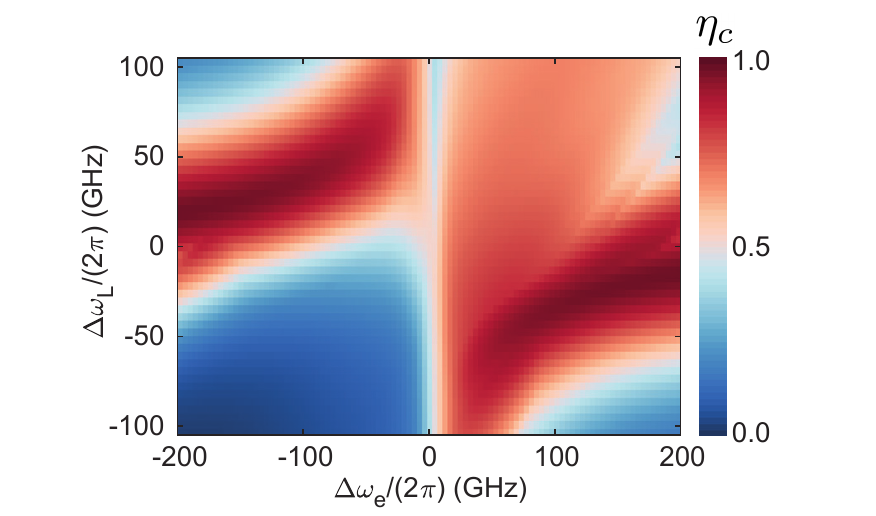}
 	\caption{$\eta_\textrm{c}$ as a function of the mode-splitting and the laser detuning. $\eta_\textrm{c}$ is above 95\% for most of the mode-splittings. Note that the bright quadrant $\Delta\omega_e>0, \Delta\omega_L>0$ is a result of phonon-mediated excitation of the QD.}
	\label{fig:MSvsLaserDet_efficiency}
\end{figure}

The success of the theory allows us to predict the behaviour on changing the mode-splitting ($\Delta\omega_e/(2\pi)$) over a large range. It is now important to calculate $\eta_\textrm{c}$ as $\beta_\textrm{c}$ depends on $\Delta\omega_e/(2\pi)$: at small $\Delta\omega_e/(2\pi)$, the two cavity modes overlap, reducing the $\beta_\textrm{c}$. Figure\,\ref{fig:MSvsLaserDet_efficiency} shows the maximum attainable $\eta_\textrm{c}$ for a range of laser detunings. The plot confirms that the maximum efficiency for a finite $\Delta\omega_e$ is achieved away from the strict resonance condition (i.e.\ $\Delta\omega_L\neq0$ and $\textrm{sign}(\Delta\omega_e)=-\textrm{sign}(\Delta\omega_L)$). The optimum laser frequency approaches the resonance of the QD as the mode-splitting increases. While near-unity efficiency is possible over a large range of mode-splittings, the photon creation efficiency shows a minimum of 50\% around $\Delta\omega_e=0$, as in this case the QD couples to both cavity modes. 

Finally, it is worth considering the input power needed for cavity-mediated excitation of a QD. For the parameters in Fig.\,\ref{fig:simulations}, the intra-cavity input pulse area for optimal efficiency is $5.4\,\pi$. This is less than the pulse area required using the phonon-assisted excitation mechanism. Furthermore, the intra-cavity field is enhanced by the high finesse of the cavity, $E_\textrm{c}=\sqrt{2F/\pi}E_\textrm{in}$ (not included in Eq.\,\ref{eq:puls_shape}). In our experiment, we used a one-sided cavity with a finesse of 500, hence giving us an enhancement of 18 in the amplitude of the field, reducing the excitation pulse area to just 0.3\,$\pi$.

\section{Conclusions}
We consider the excitation of a QD-cavity system with a laser pulse that drives the QD via a cavity mode. We show that the cavity acts as a dispersive filter, modifying the spectral configuration of the laser pulse, and that excitation of the QD proceeds via an indirect route on the Bloch sphere. Nevertheless, we demonstrate a population inversion of a QD-in-a-cavity system as high as 98\%.

Both the ``red" and ``blue" collection cases result in equivalent population inversions at the optimal parameters. In the ``red" case, the excitation mechanism resembles a rapid adiabatic passage scheme in that there is a near-unity plateau for increasing laser powers. This behaviour could be exploited to generate single photons with low sensitivity to fluctuations in the excitation power by setting the laser power to lie within the plateau. In both ``red" and ``blue" cases, the population inversion is maximum for a laser-pulse detuned with respect to the QD, illustrating the importance of a complete model of the QD-cavity system and its phonon environment to optimize the performance.

Our results demonstrate that cavity excitation of a QD can deliver the high single-photon efficiencies required for optical quantum technologies. The methods developed in this work can readily be applied to a host of different emitter and cavity systems, supporting future advances in solid-state quantum light sources. We comment on a potential drawback for the cavity-based excitation mechanism, namely that the long-lived intra-cavity field may lead to double-excitation events \cite{das2019wave,Fischer2017NPHYS}. In this case, the large-bandwidth first photon is filtered away by the cavity; the resultant time jitter on the second photon compromises the indistinguishability of the photons. This problem can be mitigated by choosing $\kappa\gg\Gamma$, where $\Gamma$ is the lifetime of the TLS in the cavity, a limit appropriate to the experiments performed here.

\section{Acknowledgments}
We acknowledge financial support from Swiss National Science Foundation project 200020\_204069, NCCR QSIT and Horizon-2020 FET-Open Project QLUSTER. A.J.\ acknowledges support from the European Unions Horizon 2020 Research and Innovation Programme under Marie Skłodowska-Curie grant agreement No.\ 840453 (HiFig), and Research Fund of the University of Basel. AJB gratefully acknowledges support from the EPSRC (UK) Quantum Technology Fellowship EP/W027909/1. S.R.V., R.S., A.L.\ and A.D.W.\ gratefully acknowledge support from DFH/UFA CDFA05-06, DFG TRR160, DFG project 383065199 and BMBF Q.Link.X.

\bibliography{bibliography_excit}

\clearpage

\onecolumngrid
\section{Supplementary}

\subsection{Normalized photon rate as a function of cavity detuning}

\begin{figure*}[ht!]
      \includegraphics[width=\textwidth]{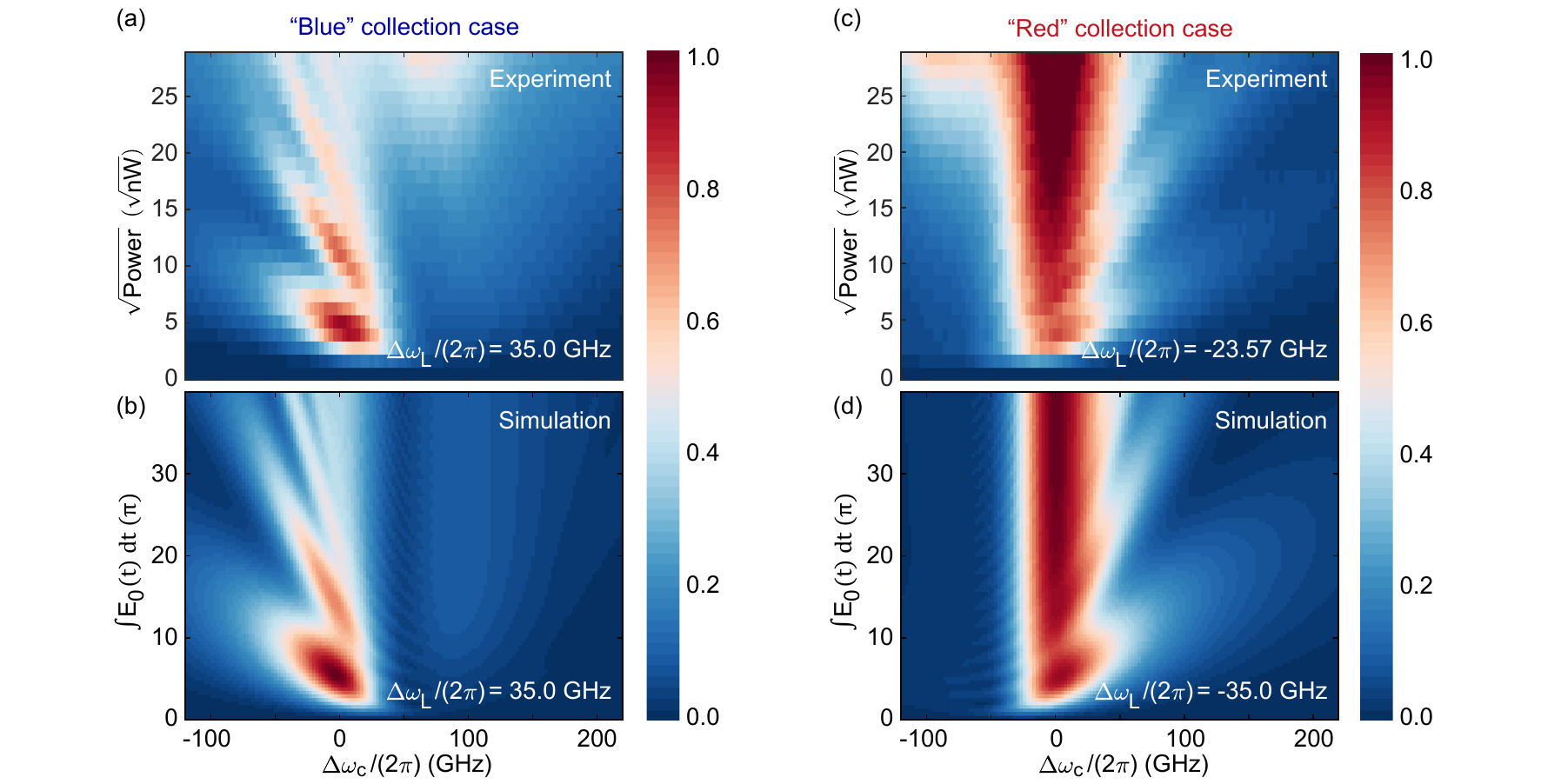}
	\caption{Normalized photon rate from a QD excited via a detuned, filtered pulse. The graphs present the photon rate as a function of the square-root of the input laser power (experimental data, top plots) or pulse area (simulations, bottom plots) and cavity detuning $\Delta\omega_\text{c}/(2\pi)$ from the QD's resonant frequency. Experimental data and simulations are scaled according to the known losses in the experimental setup. (a) Experimental data in the ``blue case", when the QD is resonantly coupled to the higher-frequency H-polarized cavity-mode and the laser pulse's central frequency is detuned by $\Delta\omega_{\text{L}}/(2\pi)=35.00$\,GHz and $t_p$ = 4.4 ps, and (b) respective simulation of $\eta_\textrm{c}$. (c) Experimental data in the ``red case", when the QD is resonantly coupled to the lower-frequency V-polarized cavity-mode and $\Delta\omega_\text{L}/(2\pi)=-23.57$\,GHz, and (d) respective simulation of $\eta_\textrm{c}$, with $\Delta\omega_{\text{L}}/(2\pi)=-35.00$\,GHz and $t_p$ = 3.6 ps.}
	\label{fig:cavdetuning}
\end{figure*}

We show in Fig.\,\ref{fig:cavdetuning} that we achieve good agreement between the experiment and the theory for the normalized photon rates as a function of cavity detuning $\Delta\omega_{\text{c}}/(2\pi)$ and excitation pulse area $\int E_0(t) dt$. In Fig.\,\ref{fig:cavdetuning}b we present the calculation in the ``blue" case when the QD is in resonance with the higher-frequency H-polarized cavity mode, and $\Delta\omega_{\rm L}/(2\pi) = 35$\,GHz and $t_p$ = 4.4 ps. At resonance, very clear damping of the Rabi rotations can be observed as a function of increasing pulse area, in very good agreement with the experimental results presented in Fig.\,\ref{fig:cavdetuning}a. A very different profile is observed in the opposite ``red" case, when the QD emits photons into the lower-frequency V-polarized cavity mode, as seen in Fig.\,\ref{fig:cavdetuning}d, where the laser is now red-detuned by $\Delta\omega_{\rm L}/(2\pi) = -35$\,GHz and $t_p$ = 3.6 ps. Here, at resonance, the normalized photon rate tends to plateau as the excitation pulse area increases, as observed in the experimental results presented in Fig\,\ref{fig:cavdetuning}c (experimental detuning $\Delta\omega_{\rm L}/(2\pi) = -23.57$\,GHz). Both the model and the experimental data show a clear asymmetry between excitation via a red- or blue-detuned cavity mode. This asymmetry is in agreement with the data presented in Fig.\,\ref{fig:Pi_res}.

\subsection{$\eta_\textrm{c}$ versus the mode-splitting of the cavity}
\begin{figure*}[ht!]
\centering
        \includegraphics[width=\textwidth]{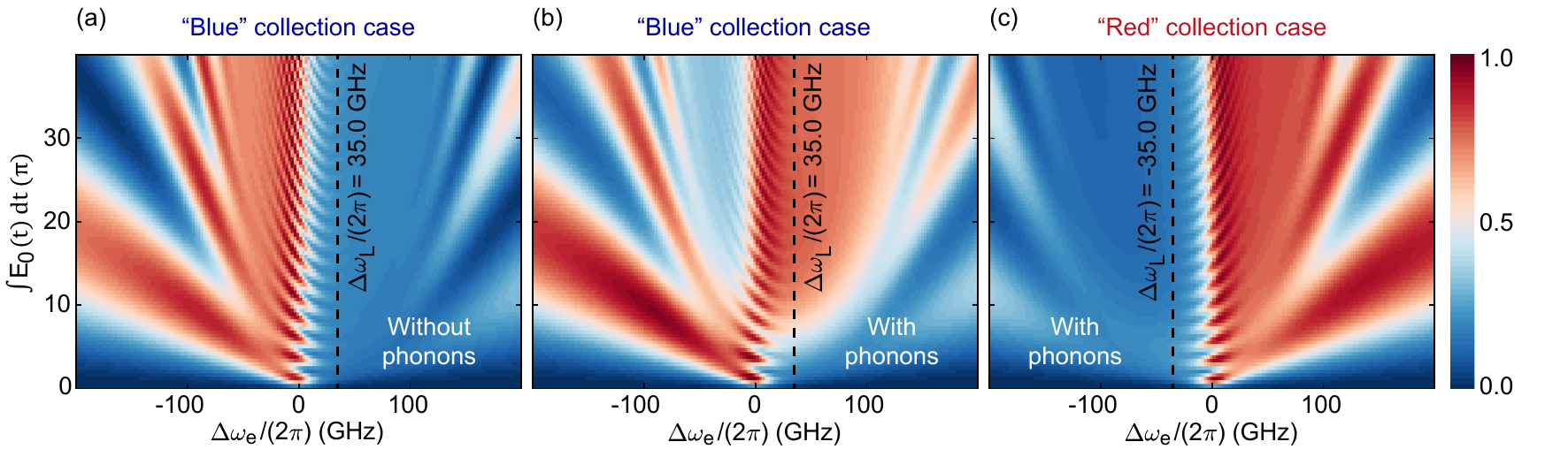}
 	\caption{$\pi_{e}$ as a function of the detuning between the excitation cavity-mode and the QD, $\Delta\omega_{\text{e}}$, and the excitation pulse area. For these simulations, $\kappa/(2\pi) = 25$\,GHz and the laser frequency is detuned from the collection cavity-mode by a fixed amount $\Delta\omega_{\text{L}}$. (a) The ``blue" collection case in the absence of phonons with  $\Delta\omega_\text{L}/(2\pi)=35.00$\,GHz and $t_p$ = 4.4 ps. (b) As in (a) but in the presence of phonons. (c) Calculated values for the ``red" collection case with $\Delta\omega_{\text{L}}/(2\pi)=-35.00$\,GHz and $t_p$ = 3.6\,ps in the presence of phonons. The dashed lines in these plots indicate the frequency of the excitation laser.}
	\label{fig:MSdetuning}
\end{figure*}

We map out the $\pi_{e}$ as a function of the frequency splitting between the excitation and the collection cavities.  Figure\,\ref{fig:MSdetuning}a shows the response of an ideal QD (no phonons) as a function of the amplitude of the input field and the mode-splitting of the cavity for a QD-laser detuning of $\Delta\omega_{\rm L}/(2\pi) = 35$\,GHz. High $\pi_{e}$ is possible for two laser detunings, the range around zero, and the range around $-60$\,GHz. The first range is disadvantageous as the QD will couple to cavity modes such that a significant fraction of the photons will be emitted into the excitation cavity. The range around $-60$\,GHz is more useful, as the detuning between the excitation cavity and the QD reduces the coupling strength to the excitation cavity. Figures \,\ref{fig:MSdetuning}b and c show $\pi_{e}$  for a QD taking into account the effect of phonons for the ``blue" and ``red" cases. As evident from these plots, $\pi_{e}$ is a weak function of the choice, ``red" or ``blue", and a high photon creation efficiency of the QD is possible for a range of laser detunings.

\end{document}